\documentstyle[11pt]{article}

\catcode`\@=11
\long\def\@makefntext#1{
\protect\noindent \hbox to 3.2pt {\hskip-.9pt
$^{{\ninerm\@thefnmark}}$\hfil}#1\hfill}		%CAN BE USED

\def\@makefnmark{\hbox to 0pt{$^{\@thefnmark}$\hss}}  %ORIGINAL

\def\ps@myheadings{\let\@mkboth\@gobbletwo
\def\@oddhead{\hbox{}
\rightmark\hfil\ninerm\thepage}
\def\@oddfoot{}\def\@evenhead{\ninerm\thepage\hfil
\leftmark\hbox{}}\def\@evenfoot{}
\def\sectionmark##1{}\def\subsectionmark##1{}}

\renewcommand{\thefootnote}{\fnsymbol{footnote}}

\newcounter{sectionc}\newcounter{subsectionc}\newcounter{subsubsectionc}
\renewcommand{\section}[1] {\vspace*{0.6cm}\addtocounter{sectionc}{1}
\setcounter{subsectionc}{0}\setcounter{subsubsectionc}{0}\noindent
	{\normalsize\bf\thesectionc. #1}\par\vspace*{0.4cm}}
\renewcommand{\subsection}[1] {\vspace*{0.6cm}\addtocounter{subsectionc}{1}
	\setcounter{subsubsectionc}{0}\noindent
	{\normalsize\it\thesectionc.\thesubsectionc. #1}\par\vspace*{0.4cm}}
\renewcommand{\subsubsection}[1]
{\vspace*{0.6cm}\addtocounter{subsubsectionc}{1}
	\noindent {\normalsize\rm\thesectionc.\thesubsectionc.\thesubsubsectionc.
	#1}\par\vspace*{0.4cm}}

\def\abstracts#1{{

\centering{\begin{minipage}{12.2truecm}\footnotesize\baselineskip=12pt\noindent
	\centerline{\footnotesize ABSTRACT}\vspace*{0.3cm}
	\parindent=0pt #1
	\end{minipage}}\par}}

\renewenvironment{thebibliography}[1]
	{\begin{list}{\arabic{enumi}.}
	{\usecounter{enumi}\setlength{\parsep}{0pt}
\setlength{\leftmargin 1.25cm}{\rightmargin 0pt}
	 \setlength{\itemsep}{0pt} \settowidth
	{\labelwidth}{#1.}\sloppy}}{\end{list}}

\newcounter{itemlistc}
\newcounter{romanlistc}
\newcounter{alphlistc}
\newcounter{arabiclistc}

 1
 1
 1

\font\ninerm=cmr9

\newcommand{\Z}{{Z \!\!\! Z}}
\newcommand{\dual}{\mbox{}^{\ast}}
\newcommand{\LL}{{I\!\! L}}
\newcommand{\half}{\frac{1}{2}}
\newcommand{\vx}{{\bf x}}
\newcommand{\vy}{{\bf y}}
\newcommand{\eq}[1]{(\ref{#1})}
\newcommand{\diff}{\partial}
\newcommand{\beq}{\begin{equation}}
\newcommand{\eeq}{\end{equation}}
\newcommand{\beqn}{\begin{eqnarray}}
\newcommand{\eeqn}{\end{eqnarray}}
\newcommand{\ddd}{{\rm d}}
\newcommand{\cD}{{\cal D}}

\def\cF{{\cal F}}
\def\cC{{\cal C}}
\def\cW{{\cal W}}
\newcommand{\cZ}{{\cal Z}}
\newcommand{\cS}{{\cal S}}

\newcommand{\intpiD}{\int\limits_{-\pi}^{+\pi} {\cD}}

\newcommand{\intinfD}{\int\limits_{-\infty}^{+\infty} {\cD}}
\newcommand{\expb}[1]{\exp\left\{ #1 \right\} }
\newcommand{\CK}[1]{\mbox{\scriptsize c}_{\mbox{$\scriptstyle #1$}}}
\newcommand{\nsum}[2]{\sum_{ #1(\CK{#2}) \in \Z }}

\newcommand{\nnsum}[2]{\sum_{\dual #1(\dual\CK{#2})
\in \Z} }
\newcommand{\nddsum}[2]{\sum_{\stackrel{\scriptstyle \dual #1(\dual\CK{#2})
\in \Z} {\delta \dual #1=0}}}
\newcommand{\dd}{\mbox{d}}
\newcommand{\fone}{\sum_{j \in \Z} \exp\left\{- \frac{4 \pi^2
\beta}{N^2} j^2 + 2 \pi i \frac{M}{N} j \right\}}
\newcommand{\ftwo}{\sum_{j \in \Z} \exp\left\{- \frac{4 \pi^2
\beta}{N^2} j^2 \right\}}

\def\dd{{\rm d}}

\def\NP{ Nucl.~Phys.}
\def\PL{ Phys.~Lett.}
\def\PR{ Phys.~Rev.}
\def\PRL{ Phys.~Rev.~Lett.}

\date{}

\begin{document}
%\vspace{-0.8cm}
\rightline{\normalsize ITEP--TH--8/95}
\rightline{\normalsize hep--th/9510014}
\vspace{1cm}
\centerline{\normalsize\bf AHARONOV--BOHM EFFECT}
\baselineskip=22pt
\centerline{\normalsize\bf IN THE ABELIAN HIGGS THEORY\footnote{
Talk given at the `Non-perturbative approaches to QCD'
workshop at ECT* in Trento.}
}
%\vspace*{0.6cm}
\vspace*{0.3cm}
\centerline{\footnotesize M.N.~CHERNODUB}
\baselineskip=13pt
\centerline{\footnotesize\it ITEP, B.Cheremushkinskaya 25, Moscow,
117259, Russia}
\centerline{\footnotesize\it and}
\centerline{\footnotesize\it Moscow Institute of Physics and Technology,
Dolgoprudny, Moscow region, Russia}
\vspace*{0.3cm}
\centerline{\footnotesize and}
\vspace*{0.3cm}
\centerline{\footnotesize M.I.~POLIKARPOV}
\baselineskip=13pt
\centerline{\footnotesize\it ITEP, B.Cheremushkinskaya 25, Moscow,
117259, Russia}
\baselineskip=12pt
\centerline{\footnotesize E-mail: polykarp@vxdesy.desy.de}
%\vfill
\vspace*{0.4cm}
\abstracts{
We study a field--theoretical analogue of the Aharonov--Bohm effect in the
Abelian Higgs Model: the corresponding topological interaction is
proportional to the linking number of the Abrikosov--Nielsen--Olesen string
world sheets and the particle world trajectory.  The creation operators of
the strings are explicitly constructed in the path integral and in the
Hamiltonian formulation of the theory. We show that the Aharonov--Bohm
effect gives rise to several nontrivial commutation relations.
We also study the Aharonov--Bohm effect in the lattice formulation of the
Abelian Higgs Model. It occurs that this effect gives rise to a nontrivial
interaction of tested charged particles.
}

\vspace*{0.4cm}
\normalsize\baselineskip=15pt
\setcounter{footnote}{0}
\renewcommand{\thefootnote}{\alph{footnote}}

\section{Introduction}

In this talk we show that the Aharonov--Bohm effect \cite{AhBo59} is
responsible for nontrivial effects in quantum Abelian Higgs theory.

It is well known that the four--dimensional Abelian Higgs theory obeys the
topologically stable classical solutions called Abrikosov--Nielsen--Olesen
(ANO) \cite{AbNiOl} strings. These strings carry the quantized magnetic
flux, wave function of charged particle which scatters the ANO string
acquires additional phase.  This shift of the phase can be measured and this
is the physical effect, which is the analogue of the Aharonov--Bohm effect:
the ANO strings playing the role of solenoids which scatter the charged
particles.  The topological long--range interaction of the Aharonov--Bohm
type between the strings and charged particles was discussed in
Refs.~\cite{AlWi89,PrKr90,KrWi89,BaMoPr93}, the same effect in the string
representation of the $4D$ Abelian Higgs model was considered on the lattice
and in the continuum limit in Refs.~\cite{PoWiZu93,AkChPoZu95}. The problem
of ANO string quantization was considered is Refs.~\cite{AkChPoZu95,Emil95}.

The Aharonov--Bohm effect in the three-- and two--dimensional Abelian Higgs
models is also interesting. The topological defects in two and three
dimensions are instantons and particles (in four dimensions they are
strings). The origin of the Aharonov--Bohm effect is the same as in the
four--dimensional case: it is also due to the topological interaction
between the particles and topological defects.

In Section~2 we repeat the calculations of Ref.~\cite{PoWiZu93} for
four--dimensional Abelian Higgs model in the continuum limit and show
explicitly the existence of the Aharonov--Bohm effect in the field theory.
We rewrite the theory in terms of the string variables and study the
interaction between the strings and particles of different charges.

Section~3 is devoted to the construction of the operator which creates the
string in a given time slice on the contour $\cC$. This operator is the
continuum analogue of the lattice operator studied in
\cite{PoWiZu93,BuPoPoWi93}.

In Section~4 we consider the theory in the Hamiltonian formalism and show
that the string creation operator has nontrivial commutation
relations\footnote{By a nontrivial commutation relation we mean a relation
of the type: $A B - e^{i\xi} B A=0$.}\, \, with the Wilson loop operator;
this is a direct consequence of the Aharonov--Bohm effect. We give several
other examples of nontrivial commutation relations.

In Section~5 for the lattice Abelian Higgs model for $D=2,3,4$ we show the
existence of the specific interaction of the test particles which is due to
Aharonov--Bohm effect. Note, that the Aharonov--Bohm effect may be also
important for the confinement problem in non--abelian gauge theories
\cite{ChPoZu94}.

\section{The Aharonov--Bohm Effect in the Abelian Higgs Model}

We discuss the four--dimensional Abelian Higgs model with the Higgs
bosons carrying the charge $N e$, the partition function can be written down
as follows:

\beq
\cZ = \int \cD A_{\mu} \cD \Phi \exp \left\{ - \int \ddd^4 x \left[
\frac{1}{4} F^2_{\mu \nu} + \frac{1}{2} |D_{\mu} \Phi|^2 + \lambda
(|\Phi|^2 - \eta^2)^2 \right] \right\}\,,
\label{Initial}
\eeq
where $D_{\mu} = \diff_{\mu} - i N e A_{\mu}$. The integration over the
complex Higgs field $\Phi = |\Phi| \exp{(i \theta)}$ can be represented as
the integration over the radial part of the Higgs field $|\Phi|$ and the
integration over the phase $\theta$. In this talk we consider the London
limit, $\lambda \rightarrow \infty$, therefore the radial part of the field
$\Phi$ is fixed and partition function \eq{Initial} can be rewritten as:

\beq
 \cZ = \int \cD A_{\mu} \cD \theta
 \exp \left\{- \int \ddd^4 x \left [ \frac{1}{4} F^2_{\mu \nu} +
 \frac{\eta^2}{2} {\left( \diff_{\mu} \theta + N e A_{\mu} \right)}^2
 \right] \right \}\,, \label{AHMN}
\eeq
where $\eta^2 = <|\Phi|^2>$.
The functional integral over $\theta$ should be carefully treated, since
$\theta$ is not defined on the manifolds where $Im \Phi = Re \Phi = 0$.
These equations define the two--dimensional surfaces which are world sheets
of the ANO strings; the Higgs field is zero at the center of the ANO string.
The variable $\theta$ can be represented as the sum of the regular and the
singular part, $\theta = \theta^r + \theta^s$. The singular part defines the
ANO string degrees of freedom. The string world sheet current
$\Sigma_{\mu\nu}$ and $\theta^s$ are related by the equations
\cite{Orl94,Lee93}:

\beqn
\diff_{[\mu,} \diff_{\nu]} \theta^s (x, \tilde x) =
2 \pi \epsilon_{\mu\nu \alpha \beta}
\Sigma_{\alpha \beta}(x, \tilde x), \label{theta}\\
\Sigma_{\alpha \beta}(x, \tilde x) =
\int_{\Sigma} \ddd \sigma_{\alpha \beta}(\tilde x) \delta^{(4)}
[x -\tilde x(\sigma)]\,, \quad
%\nonumber\\
\ddd \sigma_{\alpha \beta}(\tilde x) = \epsilon^{ab}
\diff_a \tilde x_{\alpha} \diff_b \tilde x_{\beta} \ddd^2 \sigma
\,,\nonumber
\eeqn
where $\tilde x = \tilde x(\sigma)$ are the coordinates of the
two-dimensional singularities parametrized by $\sigma_a, \ a = 1, 2$;
$\diff_a = \frac{\diff}{\diff \sigma_a}$; $\theta^s (x, \tilde x)$ is the
function of $x$ and the functional of $\tilde x$.

The integration over the fields $\theta^r$ and $A$ in eq.\eq{AHMN} leads
to the following string partition function \cite{AkChPoZu95}:

\beqn
\cZ = \int \cD \tilde x \, J(\tilde x)
\cdot \exp \left\{ - \eta^2 \pi^2 \int_{\Sigma}
\int_{\Sigma} \ddd \sigma_{\mu \nu}(\tilde x)
\cD_m^{(4)}(\tilde x - \tilde x') \ddd \sigma_{\mu \nu}(\tilde x')
\right\}, \label{ii}
\eeqn
where $\cD_m^{(4)}(x)$ is the euclidean Yukawa propagator, $(\Delta + m^2)
\cD_m^{(4)}(x) = \delta^{(4)}(x)$, and $m^2 = N^2 e^2 \eta^2$ is the mass of
the gauge boson. The action which enters the partition function \eq{ii} was
already discussed in Ref.\cite{Orl94,SaYa94}. The Jacobean $J(\tilde x)$ is
due to the change of the field variables to the string variables
\cite{AkChPoZu95}. Due to presence of the Jacobean the conformal anomaly is
absent and the quantum theory of strings \eq{ii} exists in four dimensions
(see for details the discussion in Ref's.~\cite{AkChPoZu95,Emil95}).

There exists a nontrivial long--range topological interaction of ANO strings
with particles of charge $Me$ provided the ration $\frac MN$ is
non--integer. This interaction is the four--dimensional analogue
\cite{AlWi89,PrKr90} of the Aharonov--Bohm effect studied for the lattice
Abelian Higgs model in \cite{PoWiZu93}. Here we show the existence of the
topological terms in the string representation of the theory.  Let us
consider the Wilson loop for the particle of the charge $M e$:

\beq
   W_M(\cC) = \exp\left\{ i M e \int \ddd^4 x \, j^\cC_\mu (x) A_{\mu}(x)
   \right\} = \exp\left\{ i M e \int_{\cC} \ddd x^{\mu} A_{\mu}(x)
   \right\}\,, \label{WM}
\eeq
where the current is the $\delta$--function on a contour $\cC$:

\beqn
j^{\cC}_\mu(x) = \int_{\cC} \ddd t \, \dot{\tilde z}^{\mu}(t) \,
\delta^{(4)}(x - \tilde z(t)) \label{jC}
\eeqn
and the function $\tilde z_{\mu}(t)$ parametrizes the contour.

Substituting \eq{WM} into the path integral \eq{AHMN} and changing the field
variables to the string variables we obtain:

\beqn
<W_M(\cC)> =  \nonumber \\
  \frac{1}{\cZ} \int \cD \tilde x J(\tilde x)\exp \Biggl\{
 -  \int \ddd^4 x \int
 \ddd^4 y \biggl [
 \pi^2 \eta^2 \Sigma_{\mu\nu}(x)
 \cD_m^{(4)}(x - y) \Sigma_{\mu \nu}(y)
 \nonumber \\
 + \frac{M^2 e^2}{2} j^{\cC}_\mu(x) \cD^{(4)}_m(x - y)
 j^\cC_\mu(y) + \pi i \frac{M}{N} j^\cC_\mu(x)
 \cD^{(4)}_m(x - y) \diff_{\nu} \epsilon_{\mu \nu \alpha \beta}
 \Sigma_{\alpha \beta}(y)\biggr ] \nonumber \\
 + 2 \pi i \frac{M}{N}  \LL(\Sigma,\cC)
 \Biggr\}\,, \label{bb}
\eeqn
where  $m = Ne\eta$ is the boson mass, and

\beqn
 \LL(\Sigma,\cC) = \half \int \ddd^4 x \int
 \ddd^4 y \epsilon_{\mu \nu \alpha \beta} \Sigma_{\mu\nu}(x)
 j_{\alpha}^{\cC}(y)  \diff_{\beta} \cD^{(4)}_0(x - y)  = \nonumber\\
 = \frac{1}{4 \pi^2}
 \int \dd^4 x \int \dd^4 y\, \epsilon_{\mu\nu\alpha\beta}\,
 \Sigma_{\mu\nu}(x)\, j^{\cC}_{\alpha}(y)\,\frac{{(x-y)}_{\beta}
 }{{|x-y|}^4} \label{Link4D}
\eeqn
is the linking number of the string world sheet $\Sigma$ and the trajectory
of the charged particle $\cC$, this formula represents a four--dimensional
analogue of the Gauss linking number for loops in three dimensions.
The first three terms in the exponent in \eq{bb} are short range
interactions and selfinteractions of strings and the tested particle. The
forth term is the long--range interaction which describes the
four--dimensional analogue \cite{AlWi89,PrKr90} of the
Aharonov--Bohm effect:  strings correspond to solenoids which scatter
charged particles. $\LL$ is an integer, and if $M/N$ is an integer
too, then there is no long--range interaction; this situation corresponds to
such a relation between the magnetic flux in the solenoid and the charge of
the particle when the scattering of the charged particle is absent.
The reason for this long--range interaction is that the charges $M = e,\
2e, \ \ldots (N-1)e$ cannot be completely screened by the condensate of the
field of charge $Ne$; if $M/N$ is integer, then the screening is complete
and there are no long--range forces. Note that the cloud of the screening
charges screen the Coulomb interaction between test--particles, but do
not affect the Aharonov--Bohm interaction \cite{BaMoPr93}.

Another consequence of the Aharonov--Bohm effect can be obtained, if we
consider the operator $\cF_{N}(\cS)$ \cite{KrWi89} which creates the
string with the magnetic flux $\frac{2 \pi}{N e}$ moving along a fixed
closed surface $\cS$. The operator $\cF_{N}(\cS)$ is the analogue of the
Wilson loop which creates the particle moving along the closed loop $\cC$.
An explicit form of $\cF_{N}(\cS)$ is \cite{KrWi89}:

\beq
 \cF_{N}\left(\cS\right)  =
 \exp\left\{ - \frac{\pi }{ N e}
 \int_{\cS} \dd \sigma_{\mu\nu}
  \epsilon_{\mu\nu\alpha\beta}\, F_{\alpha\beta} (x)\right\}\,.  \label{FE}
\eeq

There exists an operator which can be calculated {\it exactly},
\cite{PrKr90}; this operator is the normalized product of the Wilson
loop $W_M(\cC)$ and $\cF_{N}(\cS)$:

\beq
 A_{NM}(\cS,\cC) = \frac{ \cF_N(\cS) W_M(\cC)}{<\cF_N(\cS)> <W_M(\cC)>}.
\eeq
Here $<\cF_{N}(\cS)>$ is a constant which depends on the regularization
scheme. Substituting this operator into the functional integral \eq{AHMN}
and integrating over the fields $A$ and $\theta$, we obtain the following
result:

\beq
  <A_{NM}(\cS,\cC)> = e^{2 \pi i \frac{M}{N} \LL(\cS,\cC)}. \label{AMN}
\eeq
The meaning of this result is very simple. If the surface $\cS$ lies in a
given time slice, then $<A_{NM}(\cS,\cC)> = \exp\left\{\frac{2\pi
i}{Ne}Q_{\cS}\right\}$ (see \cite{PrKr90,KrWi89}), where $Q_\cS$ is
the total charge inside the volume bounded by the surface $\cS$; if
$\LL(\cS,\cC) = n$, then there is the charge $Mne$ in the volume bounded by
$\cS$.

\newpage

\section{The String Creation Operator.}

In the previous Section we have derived the partition function of the
Abelian Higgs model as a sum over the closed world sheets of the ANO
strings. Now we construct the operator which creates the string on a closed
loop at a given time. The vacuum expectation value of this operator is the
sum over all surfaces spanned on a given loop.  A similar operator for the
lattice theory was suggested in \cite{BuPoPoWi93,PoWiZu93}. The construction
is similar to the soliton creation operator suggested by Fr\"{o}hlich and
Marchetti \cite{FrMa87}. First we consider the model \cite{Lee93} which is
dual to the original Abelian Higgs model. It contains the gauge field
$B_\mu$ dual to the gauge field $A_\mu$, and the hypergauge field
$h_{\mu\nu}$ dual to $\theta^r$. In order to get it we change the
integration in $\theta^s$ to the integration over $\tilde x$ and make the
duality transformation, the details of this transformation are given in
\cite{Lee93}. Taking into the account the Jacobian \cite{AkChPoZu95}, we
get:

\beqn
\cZ =  \int \cD h \cD B \cD \tilde x J(\tilde x)
\exp \biggl\{ - \int \ddd^4 x \biggl[
\frac{1}{3 \eta^2} H^2_{\mu \nu \alpha} + \frac{e^2 N^2}{2} (h_{\mu
\nu} - \diff_{\mu} B_{\nu} + \diff_{\nu} B_{\mu})^2 \nonumber \samepage \\
+ 2 \pi i h_{\mu\nu} \Sigma_{\mu\nu}
(x, \tilde x) \biggr] \biggr\}\, , \label{Dual}
\eeqn
where $H_{\mu \nu \sigma} = \diff_\mu h_{\nu\sigma} + \diff_\nu h_{\sigma
\mu} + \diff_\sigma h_{\mu\nu} $ is the field strength of the
hypergauge field $h_{\mu\nu}$.
The action of the dual theory is invariant under the gauge
transformations:  $B_{\mu}(x)~\rightarrow~B_\mu(x) +
\diff_{\mu}\alpha(x)$, $h_{\mu\nu}(x)~\rightarrow~h_{\mu\nu}(x)$, and under
the hypergauge transformations: $B_{\mu}(x) \rightarrow B_{\mu}(x) -
\gamma_{\mu}(x)$, $h_{\mu\nu}(x) \rightarrow h_{\mu\nu}(x) +
\diff_{\mu}~\gamma_{\nu}(x) - \diff_{\nu}~\gamma_{\mu}(x)$.

The ANO string carries magnetic flux, and in order to construct the
creation operator, it is natural to use the dual Wilson loop: $\cW_D(\cC) =
\exp\{i \int \ddd^4 x B_{\mu}(x) j^\cC_\mu(x)\}$, where the current
$j^\cC_\mu(x)$ defines the loop $\cC$ \eq{jC}. This operator is gauge
invariant but it is not hypergauge invariant, and its vacuum expectation
value is zero. To construct the hypergauge invariant operator
\cite{BuPoPoWi93,PoWiZu93}, we follow an idea of Dirac \cite{Dir55}, who
suggested the gauge invariant creation operator of a particle with the
charge $M$:

\beq
\Phi^c_M(\vx) = \Phi_M(\vx) \exp \left \{i M e
\int \ddd^3 y G_{l}(\vx - \vy) A_{l}(\vy) \right \}\, , \label{Phic}
\eeq
here $\diff_{i} G_{i}(\vx) = \delta^{(3)}(\vx)$, and the gauge variation of
the matter field $\Phi(x) \rightarrow \Phi_M(x) \, e^{i M e
\alpha(x)}$ is compensated by the gauge variation of cloud of photons
$A_\mu$. Now we use a similar construction, namely, we surround $\cW_D(\cC)$
by the cloud of the Goldstone bosons:

\beq
   U(\cC) = \cW_D(\cC) \exp\left\{\frac{i}{2} \int \ddd^3 y G^{ij}_{\cC}(x -
   y) h_{i j}(y)\right\}\, . \label{Uc}
\eeq
It is easy to see that $U(\cC)$ is hypergauge invariant if the
skew--symmetrical tensor $G^{ij}_{\cC}(x)$ satisfies the
equation\footnote{In this and in the next sections, the Latin indices vary
from 1 to 3 and the Greek indices vary from 0 to 3.} $\diff_{i}
G^{ik}_{\cC}(x) = j^{\cC}_k(x)$.  It is convenient to choose
$G^{ik}_{\cC}(x)$ as the surface, spanned on the loop $\cC$:  $G^{ij}_{\cC}
= \int_{\cS_\cC} \ddd \sigma^{ij}(\tilde x) \delta^{(4)} [x - \tilde
x(\sigma)]$ ({\it cf.} eq.\eq{theta}). Since the string creation operator
should act at a definite time slice, the surface defined by
$G^{ij}_{\cC}(x)$ and the loop $\cC$ should belong to that time
slice\footnote{The solution of the equation $\diff_{i} G^{ik}_{\cC}(x) =
j^{\cC}_k(x)$ is non--unique, moreover we choose a two dimensional surface
as the support of $G^{ik}$, the solution which has three--dimensional
support can be of the form: $G^{ik}_{\cC} = \int \ddd^3 y \diff^{[i}
j^{k]}_{\cC}(\vy) \cD^{(3)}_0(\vx - \vy) $, where $\cD^{(3)}_0 = -
\frac{1}{4 \pi |\vx-\vy|}$. It is easy to find that all these ambiguities do
not change the physical results.}.

Substituting the operator \eq{Uc} into the dual partition function \eq{Dual}
and performing the inverse duality transformation, we get the vacuum
expectation value of the string creation operator in terms of the original
fields $A$ and $\theta$:

\beqn
    < U(\cC)> = \frac{1}{\cZ} \int \cD A \cD \theta
    \exp\biggl\{ - \int \ddd^4 x \biggl[ \frac{1}{4} {\left( F_{\mu\nu} +
    \frac{2 \pi}{N e} \epsilon_{\mu\nu\lambda\sigma} G^{\lambda
    \sigma}_{\cC}(x) \right)}^2
    \nonumber\\
    + \frac{\eta^2}{2} {\left( \diff_{\mu} \theta + N e A_{\mu} \right)}^2
    \biggr] \biggr\}\, , \label{UCor}
\eeqn
where the tensor $G^{\mu\nu}_{\cC}$ is equal to $G^{ij}_{\cC}$ if $\mu=i$
and $\nu=j$ are spatial indices, and $G^{0\mu}_{\cC}= G^{\mu0}_{\cC} =0$ for
any $\mu$. If we change the field variables in \eq{UCor} to the string
variables, we get a sum over the closed surfaces $\Sigma$:

\beqn
<U(\cC)> = \frac{1}{\cZ} \int \cD \tilde x J(\tilde x)
\exp \left \{- \eta^2 \pi^2 \int \ddd^4 x \int \ddd^4 y \left [
\left(\Sigma^{\mu \nu}(x, \tilde x) +
\right. \right. \right. \label{<Uc>} \\
\left. \left. \left. G^{\mu \nu}_{\cC}(x) \right)
\cD^{(4)}_m(x - y) \left(\Sigma^{\mu \nu}(y, \tilde x) + G^{\mu
\nu}_{\cC}(y) \right) \right ] \right \}\,, \nonumber
\eeqn
where $\Sigma^{\mu \nu}(x, \tilde x)$ is defined by eq.\eq{theta}.

The summation over all closed surfaces $\Sigma^{\mu \nu}$, plus the open
surface $G^{\mu \nu}$ with the boundary $\cC$, is equivalent to the
summation over all closed surfaces and over {\it all} surfaces spanned on
the loop $\cC$. Therefore, the operator $U(\cC)$ creates a string on the
loop $\cC$. Using the string creation operators, it is easy to construct the
operators which correspond to the processes of decay and scattering of the
strings.

\section{Aharonov--Bohm Effect In The Hamiltonian Formalism.}

In this section we consider the ANO strings in the framework of the
canonical quantization. We start with the standard commutation relations:
$\left[\pi^i\left(\vx\right)\,,\, A^j\left(\vy\right)\right]= - i \delta_{ij}
\delta\left(\vx-\vy\right)$, $\pi^{i} = F^{0i}$ and
$\left[\pi_{\phi}\left(\vx\right)\,,\, \phi\left(\vy\right) \right] = - i
\delta\left(\vx - \vy\right)$, $\pi_{\phi} =
{\left(D^0\phi\right)}^*$.  Using the string creation operators \eq{FE} and
\eq{Uc}, we construct several operators, which satisfy the commutator
relations of the type:  $A\cdot B - B \cdot A e^{i\xi} = 0$. Similar
operators are known for $3D$ Abelian models, see for example
refs.\cite{D3}. The physical phenomenon leading to the nontrivial
commutation relations in the nonabelian theories was discussed by
't~Hooft~\cite{tHo78}.

First, let us consider the operator $U_{str}(\cC)$ which creates the
ANO string on the loop $\cC$:

\beq
   U_{str}(\cC) = \exp\left\{ \frac{2 \pi i}{N e} \int \ddd^3 x \half
   \epsilon_{ijk} G^{ij}_{\cC}(x) \pi^{k}(x) \right\}\,, \label{Ustr}
\eeq
here $G^{ij}_{\cC}(x)$ is the same function as in
eq.\eq{Uc}. The operator \eq{Ustr} is a special case of the creation
operator:

\beq
   U[A^{cl}] = \exp\left\{ i \int \ddd^3 x
   A^{cl}_k (x) \pi^{k}(x) \right\}\,, \label{UstrA}
\eeq
where $A^{cl}(x)$ is a classical field. It is easy to see, that
$U[A^{cl}] |A(x)> = |A(x) + A^{cl}(x)>$. In \eq{Ustr} we have $A^{cl}_k (x) =
\frac{2 \pi}{N e} \epsilon_{ijk} G^{ij}_{\cC}(x)$, and the
magnetic field corresponds to the infinitely thin string on the loop $\cC$:
$B_i(x) =  \frac{2 \pi}{N e}\,  j^{\cC}_i(x)$; the current
$j_i^{\cC}$ is defined by eq. \eq{jC}.

The commutation relations for the operator \eq{Ustr} with the operators of
the electric charge $Q = \int \dd^3 x \diff_i \pi^i(\vx)$ and the magnetic
flux $\Phi_{i} = \int \dd^3 x \epsilon_{ijk} \diff^{j} A^{k}(\vx)$ also
show that $U_{str}(C)$ creates a string which carries the magnetic flux
$\frac{2 \pi}{N e}$ on the contour $\cC$:

\beqn
  \left[Q\left(x_0,\vx\right)\,,\,
  U_{str}(\cC) \right] = 0\,, \quad
  \left[\Phi^j\left(x_0,\vx\right)\,,\,
  U_{str}(\cC) \right] = \frac{2 \pi}{N e}
  j^{\cC}_i(x) U_{str}(\cC)\,. \label{commUc}
\eeqn

Note that, the string creation operator \eq{UCor} considered in the
previous section can be rewritten in the following way:

\beq
U(\cC) =
\exp\left\{ - \int \ddd^4 x \left[ \frac{1}{4} {\left( F_{\mu\nu} +
\frac{2\pi}{N e} \epsilon_{\mu\nu\lambda\sigma}
G^{\lambda \sigma}_{\cC}(x)\right)}^2 -
\frac{1}{4} F_{\mu\nu}^2\right] \right\},
\eeq
and it is clear
that, up to an inessential factor, it coincides with \eq{Ustr}.

Now we consider the commutator of the operator $U_{str}(\cC_1)$ and the
Wilson loop $W_M(\cC_2)$ \eq{WM}, the contours $\cC_1$ and $\cC_2$ belong to
the same time slice. Using the relation $e^A e^B = e^B e^A e^{[A,B]}$, which
is valid if $[A,B]$ is a $c$--number, it is easy to get:

\beq
  U_{str}(\cC_1) W_M(\cC_2) - e^{i \xi(\cC_1,\cC_2)} W_M(\cC_2)
  U_{str}(\cC_1) = 0\,, \label{WUc}
\eeq
where
$\xi(\cC_1,\cC_2) = \frac{2 \pi M}{N} \LL(\cC_1,\cC_2)$, and
$\LL(\cC_1, \cC_2) = \frac{1}{4 \pi} \int_{\cC_1} \ddd x_i \int_{\cC_2} \ddd
y_i \epsilon^{ijk} \frac{{(x-y)}_k}{|\vx-\vy|^3}$ is the linking number of
the loops $\cC_1$ and $\cC_2$, see Fig.1. The commutation relation
\eq{WUc} is the direct consequence of the Aharonov--Bohm effect; the wave
function of the particle of the charge $M e$ acquires the additional phase
$e^{\frac{2 \pi i M}{N}}$ if it goes around a solenoid with the magnetic
flux $\frac{2 \pi}{N e}$.

The next example is the commutation relation of the Dirac operator
$\Phi^c_M(\vx)$ \eq{Phic} which creates the particle with charge $M$ at the
point $\vx$, and the operator $\cF_N(\cS)$  which creates the string on the
surface $\cS$. In Minkowsky space, the operator $\cF_N(\cS)$ has the form
(an analogue of eq.\eq{FE}):

\beq
 \cF_{N}\left(\cS\right)  =
 \exp\left\{ \frac{ i \pi }{ N e}
 \int_{\cS} \dd \sigma_{\mu\nu}
  \epsilon_{\mu\nu\alpha\beta}\, F_{\alpha\beta} (x)\right\}\,.  \label{F}
\eeq
If the surface $\cS$ belongs to the same time slice as the point $\vx$,
then:

\beq
   \cF_N(\cS) \Phi^c_M(\vx) - \Phi^c_M(\vx) \cF_N(\cS)
   e^{i\theta(\cS,\vx)} = 0\,, \label{FPh}
\eeq
where $\theta\left(\cS,\vx\right) = \frac{2 \pi M}{N}
\Theta(\cS,\vx)$.  The function $\Theta(\cS,\vx)$ is the "linking
number" of the surface $\cS$ and the point $\vx$, see Fig.2:

\beq
  \Theta\left(\cS,\vx \right) = \cases{1& if $x$ lies inside
  volume bounded by $\cS$; \cr 0& otherwise}\,. \label{Theta}
\eeq
It is obvious that the commutation relation \eq{FPh} is also a consequence
of the Aharonov--Bohm effect.

Consider now the composite operator
\beq
  H_{MN}(\vx,\cS) = \Phi^c_M(\vx)\,
  \cF_N\left(\cS\right)\,,\label{H}
\eeq
where the surface $\cS$ lies at the same time slice as the
point $\vx$.
Using commutation relation \eq{FPh} it is easy to find that:

\beqn
   H_{M_1,N}\left(\vx_1,\cS_1 \right) H_{M_2,N}\left(\vx_2,\cS_2
   \right) - H_{M_2,N}\left(\vx_2,\cS_2\right) H_{M_1,N}
   \left(\vx_1,\cS_1 \right) e^{i \zeta_{12}}\,= 0 \, ,
   \label{comm:H}
\eeqn
where $\zeta_{12} = \frac{2 \pi M_1}{N}
\Theta\left(\vx_1, \cS_2\right) - \frac{2 \pi M_2}{N} \Theta
\left(\vx_2, \cS_1\right)$.
If the point $\vx_1$ lies in the volume bounded by $\cS_2$, the point
$\vx_2$ lies out of the volume $\cS_1$, $M_{1,2}=1$ and \ $N=2$, then
eq.\eq{comm:H} leads to the fermion--like commutation relation

\beq
  H(\vx_1) H(\vx_2) + H(\vx_2) H(\vx_1) = 0\, , \label{HH}
\eeq
where $H(\vx_i) = H_{M_i}(\vx_i, \cS_i)$.

The commutation relations \eq{comm:H} and \eq{HH} can be explained as
follows. The operator $\cF_N(\cS)$ creates the closed world sheet of the ANO
string and the configuration space of the (charged) particles becomes not
simply connected. Similar reasons lead to nontrivial statistics in $2+1$
dimensions \cite{Rao92}. Note that all operators and commutation relations
considered in the present section can be constructed in the free theory, but
the states created by the operators $U_{str}(\cC_1)$ and $\cF_N(\cS)$ are
unstable in this case. In the Abelian Higgs theory, the ANO strings exist as
a solution of the classical equations of motion, and this fact justifies the
study of the commutation relations which contain string creation operators.

\section{Aharonov--Bohm Effect on the Lattice}

Now we study the Aharonov--Bohm effect on the lattice. For the sake of
convenience we use the differential forms formalism on the lattice (see
Ref.\cite{Intro} for the introduction). As in the previous Sections we
consider the Abelian Higgs model in the London limit. The corresponding
partition function in the Villain form is given by the following formula:

\beq
  \cZ = \intinfD A \intpiD \varphi \nsum{l}{1} \expb{
  - \beta \|\dd A\|^2
  - \gamma \| \dd\varphi + 2\pi l - N A \|^2}\,, \label{NC}\\
\eeq
where $A$ is the noncompact gauge field, $\phi$ is the phase of the Higgs
field, $l$ is the integer--valued one--form and we do not specify the
dimension $D$ of the space--time. In order to rewrite the partition function
\eq{NC} in terms of topological defects we should perform the so called
Berezinski--Kosterlitz--Thauless (BKT) transformation \cite{BKT} with
respect to the form $l$. First change the summation variable,
$\displaystyle{\nsum{l}{1} = \sum_{\stackrel{\scriptstyle \sigma(\CK{2}) \in
\Z} {\dd \sigma=0}} \nsum{q}{0}}$, here $l = m[\sigma] + \dd q$ and
$m[\sigma]$ is a particular solution of the equation $\dd m[\sigma] =
\sigma$. Using the Hodge decomposition $m[\sigma] = \delta \Delta^{-1}
\sigma + \dd \Delta^{-1} \delta m[\sigma]$ we introduce the noncompact field
$\varphi_{n.c.} = \varphi + 2 \pi (\Delta^{-1}\delta m[\sigma] +q), \
\displaystyle{ \nsum{q}{0}} \intpiD \varphi = \intinfD \varphi_{n.c}$, the
integration over the noncompact fields $A$ and $\varphi_{n.c.}$ gives:

\beq
  \cZ^{BKT} = const. \, \nddsum{\sigma}{D-2} \expb{ - 4 \pi^2 \gamma
  \left(\sigma, {(\Delta + m^2)}^{-1} \sigma \right)} \,, \label{TD0}
\eeq
where $m^2 = N^2 \gamma \slash \beta$ is the mass of the vector boson field.
The closed $(D-2)$--forms $\sigma$ represents the world paths for the
topological defects of the theory, which are instanton--like defects in
$D=2$, particle--like defects in $D=3$ and ANO strings in $D=4$ (compare the
lattice equation \eq{TD0} for $D=4$ with eq.\eq{ii} in the continuum).

The same transformations applied to the quantum average of the Wilson loop
$W_M(\cC) = \exp\{ i M (A,j_\cC) \}$ leads to the following formula:

\beqn
 <W_M(\cC)> = \frac{1}{\cZ^{BKT}}\nddsum{\sigma}{D-2} \expb{-
 4\pi^2\gamma \left(\dual \sigma,{\left(\Delta + m^2\right)}^{-1}\dual
 \sigma \right) \right.  \makebox[5.4em]{} \nonumber\\
 \left.- \frac{M^2}{4 \gamma}( j_\cC,(\Delta + m^2)^{-1} j_\cC) - 2 \pi i
 \frac{M}{N} \left(j_\cC,{\left(\Delta + m^2\right)}^{-1}\delta
 \sigma \right) + 2 \pi i \frac{M}{N} \LL\left(\sigma,j_\cC\right)}\,,
 \label{wl}
\eeqn
The first three terms in this expression are short--range Yukawa forces
between defects. The last term is long--range and it has the topological
origin: $\LL (\sigma,j_\cC)$ is the linking number between the world
trajectories of the defects $\sigma$ and the Wilson loop $j_\cC$:

\beq
       \LL(\sigma,j_\cC) = (\dual j_\cC, {\Delta}^{-1} \dd \dual \sigma)\,.
\label{Ll}
\eeq
$\LL$ is equal to the number of points at which the loop $j_\cC$ intersects
the $(D-1)$--dimensional volume $\dual n$ bounded by the closed
$(D-2)$--manifold defined by the form $\dual \sigma(\dual c_{D-2})$,
see \cite{PoWiZu93} for $D=4$. For four dimensions eq.\eq{Ll} is the lattice
analogue of eq.\eq{Link4D}. Note that eq.\eq{wl} for the quantum average of
the Wilson loop remains unchanged under the transformation\footnote{This is
valid only for the Abelian Higgs Model with the noncompact gauge field
$A$.} \, \, $N \to 1$, $M \to \frac MN$, $\beta \to \frac{\beta}{N^2}$.

Let us first consider the quantum average \eq{wl} in two dimensions. In the
limit $\gamma \gg \beta$ (the vector boson mass in the lattice units is much
greater than unity) this equation reduces to

\beq
 <W_M(\cC)> = \frac{1}{\cZ^{BKT}}\nddsum{k}{0} \expb{-
  \frac{4 \pi^2\beta}{N^2} {\| \dual k \|}^2 +
 2 \pi i \frac{M}{N} \LL\left(k,j_\cC\right)}\,,
\eeq
where the zero--form $\dual k$ represents the instanton "world paths" and we
omitted the terms of the order $O(m^{-2})$.

The linking number can be rewritten as $\LL(k,j_\cC) = (k, m_\cC)$, where
two--form $m_\cC$ represents the surface spanned on the contour $\cC$, and
the result is:

\beq
 <W_M(\cC)> ={\left( \frac{\fone}{\ftwo} \right)}^{\cS(\cC)}
 = const. \exp\left\{ - \sigma(M, N;\gamma) \cdot \cS(\cC) \right\}\,,
 \label{D2W}
\eeq
where $\cS(\cC)$ is the area of the surface which is spanned on the contour
$\cC$.

  The string tension $\sigma(M, N;\gamma)$ is given by the equation:

\beq
 \sigma(M, N;\gamma) = - ln\left[ \frac{\Theta\left(
 \frac{4 \pi^2\beta}{N^2}, \frac{M}{N} \right)}{\Theta\left( \frac{4
 \pi^2\beta}{N^2},0\right)}\right]\,,
 \label{sigma1}
\eeq
where
$\Theta(x,q)$ is $\Theta$--function:

\beq
    \Theta(x,q) = \sum_{j \in \Z} \exp\left\{ - x j^2 + 2 \pi i q j
    \right\}\,.
    \label{Th}
\eeq
Note, that the string tension $\sigma$ has interesting property: if the
charge of the test particle is completely screened by the charge of the
Higgs condensate $\left( \frac MN \in \Z \right)$, the string tension is
zero, since $\Theta(x,q + 2 \pi n) = \Theta(x,q)$ where $n$ is integer.
The dependence of the string tension $\sigma(M, N;\gamma)$ on $\beta$ for $M
= 1$ and $N=2,3,4$ is shown on Fig.3. The area law of the Wilson loops
for fractionally charged particles in the $2D$ abelian Higgs model was first
found in Ref.~\cite{CaDaGr}, but it was not realized that the nonzero value
of the string tension is due to the analogue of the Aharonov--Bohm effect.

Now we consider the three--dimensional Abelian Higgs model in the limit
$\gamma \gg \beta$. The analogue of eq.\eq{D2W} is given by the formula:

\beq
 <W_M(\cC)> = \frac{1}{\cZ^{BKT}}\nddsum{j}{1} \expb{
 \frac{4 \pi^2\beta}{N^2} {\| \dual j \|}^2 +
 2 \pi i \frac{M}{N} (\dual m_\cC,\dual j)} \,,
 \label{A7}
\eeq
where the one--form $\dual j$ represents the world trajectories of the
vortices and $m_\cC$ is again an arbitrary surface, which is spanned on the
contour $\cC$.

Let us represent the condition of closeness of the vortex lines $j$ in
eq.\eq{A7} by the introduction of the integration over additional compact
form $C$:

\beq
  \nddsum{j}{1} \cdots = \nnsum{j}{1} \delta \left( \delta \dual j
  \right) \cdots = \intpiD \dual C \nnsum{j}{1}\, \exp
  \left\{ i (\delta \dual j, \dual C) \right\} \cdots \,.
  \label{A8}
\eeq
Inserting the unity

\beq
    1 = \intinfD \dual F \delta \left( \dual F - \dual j \right)
\eeq
in eq.\eq{A7}, we get:

\beqn
  <W_M(\cC)> = \frac{1}{\cZ^{BKT}} \intinfD \dual C \intinfD \dual F
  \nnsum{j}{1} \cdot \nonumber\\
  \exp \left\{ - \frac{4 \pi^2\beta}{N^2} {\| \dual F \|}^2
  + 2 \pi i \frac{M}{N} (\dual m_\cC,\dual F) + i (\delta \dual F, \dual C)
  \right\} \cdot \delta \left( \dual F - \dual j \right) \,.
  \label{A9}
\eeqn
The use of the Poisson formula $2 \pi \sum_{n \in \Z} \delta(n-x) = \sum_{n
\in \Z} \, e^{i n x}$ and integration out of the fields $\dual F$ leads to
the dual representation of the quantum average \eq{A7}:

\beqn
  <W_M(\cC)> = \frac{1}{\cZ^{BKT}}
  \nnsum{j}{1} \intinfD C \, \exp \left\{ - \frac{N^2}{16 \pi^2 \beta}
  {\| \dd \dual C + 2 \pi \frac{M}{N} \dual m_\cC + 2 \pi \dual j\|}^2
  \right\}\,.
  \label{A10}
\eeqn
At $\beta \to 0$ we can evaluate this expression in the semiclassical
approximation, the result is:

\beq
  <W_M(\cC)> = const. \, \exp \left\{ - \frac{q^2 N^2}{4 \beta} (j_\cC,
  \Delta^{-1} j_\cC) \right\} \,,
\eeq
where

\beqn
  q = \min_{K \in \Z} |\frac{M}{N} - K|\,.
  \label{q}
\eeqn

Let us consider the large Wilson loop of the size $T \times R$, $T \gg R \gg
1$.  Taking into account that two--dimensional massless propagator
$\Delta^{-1}(x)$ behaves at large $x$ as $\Delta^{-1}(x) = C_0 \cdot \ln x +
\dots$, where $C_0$ is the numerical constant, we find in the leading order:

\beq
  <W_M(\cC)> = const. \, \exp \left\{ - \kappa(M, N; \beta) \cdot
  T \ln R \right\}\,,
  \label{wl:kappa}
\eeq
where the coefficient $\kappa(M, N; \beta)$ is given by the formula:

\beq
    \kappa(M, N; \beta) = C_0 \, \frac{q^2 N^2}{4 \beta}\,,
    \label{kappa1}
\eeq
$q$ is defined by eq.\eq{q}.

We see, that the Aharonov--Bohm effect leads to the logarithmic potential in
$3D$ for the test particles of the charge $M e$. If $\frac{M}{N}$ is integer
(complete screening) this potential vanishes. The logarithmic
behavior of the potential was also predicted from some simple considerations
in Ref.~\cite{Sa79}.

In the four--dimensional theory the Aharonov--Bohm interaction leads to
corrections to the perimeter law ({\it i.e.} to mass renormalization).
Nevertheless this effect can be important at finite temperature. When the
temperature increases the system becomes closer to $3D$ theory and at some
temperatures the ANO strings give rise to the logarithmic term in the
potential extracted from the spatial Wilson loop. The behavior of the Wilson
loops for the lattice $4D$ Abelian Higgs model with the compact gauge field
at finite temperature was studied both analytically \cite{BaRa79} and
numerically \cite{4DTn}. In the phase diagram of compact version of the
theory the confining region exists and therefore the logarithmic behavior is
difficult to investigate.  We are planning to investigate numerically the
noncompact Abelian Higgs model at finite temperature in which the
logarithmic potential should not be suppressed.

In this section we found several formulae, eqs.(\ref{sigma1},\ref{Th}) and
eqs.(\ref{q},\ref{kappa1}), which depend on the fractional part of the ratio
$\frac{M}{N}$. This remarkable dependence is due to the Aharonov--Bohm
effect.

\vspace*{0.6cm}
\noindent
{\normalsize\bf Acknowledgments}\par\vspace*{0.4cm}

The authors are grateful to A.~Kovner, A.~Morozov and M.~de~Wild~Propitius
for interesting discussions.  This work was supported by the Grant No.
MJM000, financed by the International Science Foundation, by the Grant No.
MJM300, financed by the International Science Foundation and by the
Government of the Russian Federation, by the JSPS Program on Japan -- FSU
scientists collaboration, by the Grant INTAS-94-0840 and by the Grant No.
93-02-03609, financed by the Russian Foundation for Fundamental Sciences.

\vspace*{0.6cm}
\noindent
{\normalsize\bf Figure Captions}\par\vspace*{0.4cm}

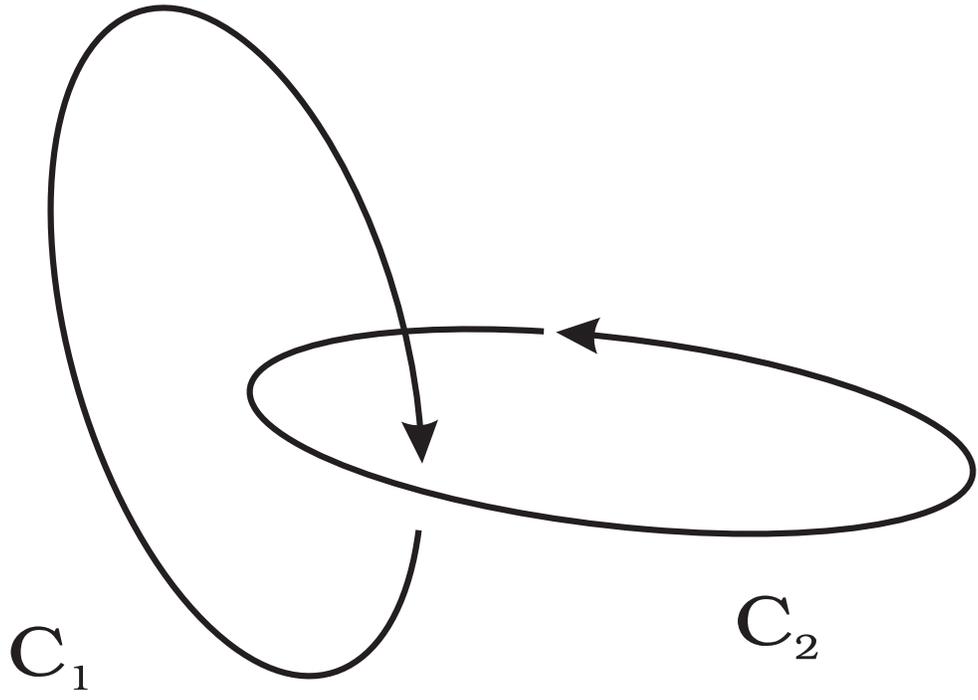
Fig.1. Linking of the contours $\cC_1$ and $\cC_2$ in $3D$;

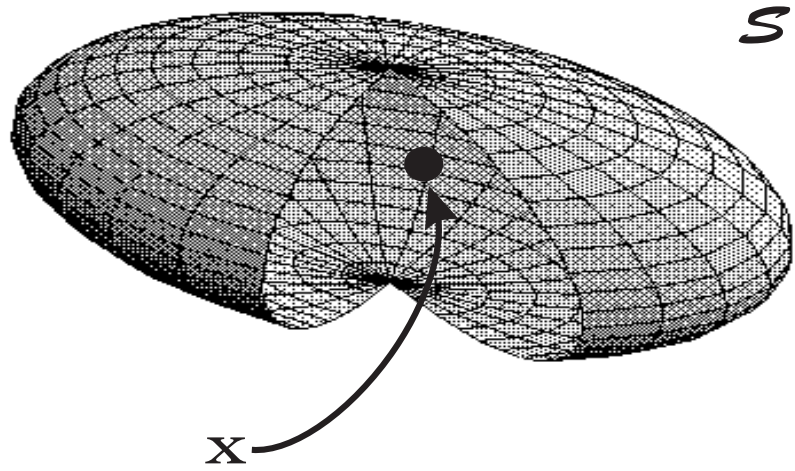
Fig.2. Linking of the surface $\cS$ and the point $\vx$ in $3D$;

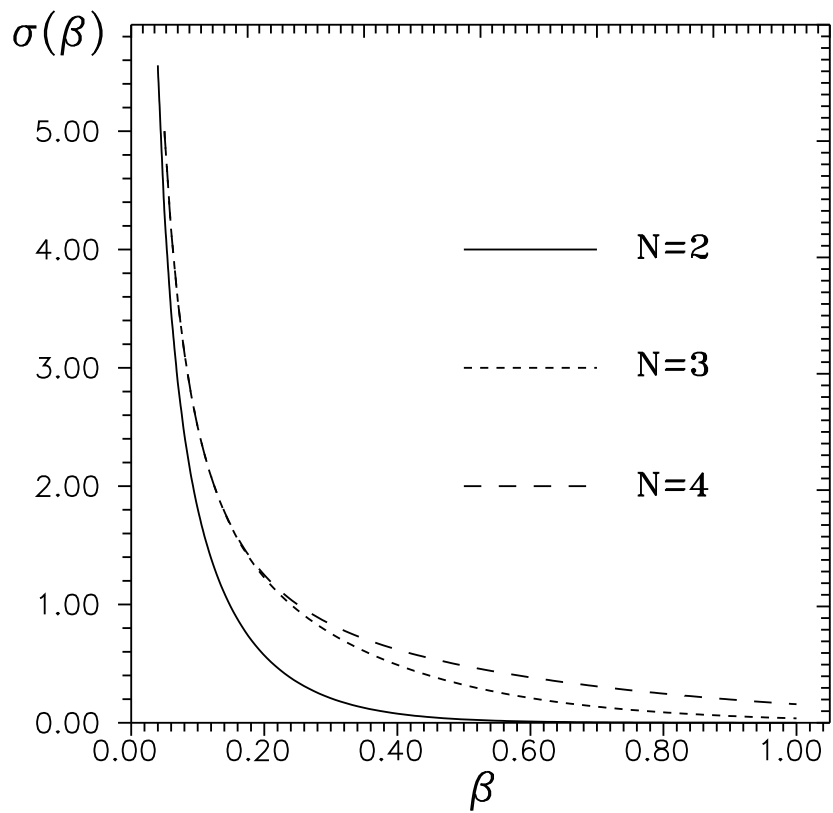
Fig.3. The dependence of the string tension $\sigma(M, N;\gamma)$
       (eq.\eq{sigma1}) on $\beta$ for $M = 1$ and

       $N=2,3,4$

\newpage
\noindent
{\normalsize\bf References}\par

\end{document}